\documentclass[fleqn,usenatbib]{mnras}

\usepackage{newtxtext,newtxmath}
\usepackage[T1]{fontenc}

\DeclareRobustCommand{\VAN}[3]{#2}
\let\VANthebibliography\thebibliography
\def\thebibliography{\DeclareRobustCommand{\VAN}[3]{##3}\VANthebibliography}

\usepackage{graphicx}	% Including figure files
\usepackage{amsmath}	% Advanced maths commands
\usepackage{subcaption}

\title[Modelling A Hot Horizon in Global 21\,cm Experimental Foregrounds]{Modelling A Hot Horizon in Global 21\,cm Experimental Foregrounds}

\author[J. H. N. Pattison et al.]{
Joe H. N. Pattison,$^{1,2}$\thanks{E-mail: jhnp2@cam.ac.uk}
Dominic J. Anstey,$^{1,2}$\thanks{E-mail: da401@cam.ac.uk}
Eloy de Lera Acedo$^{1,2}$\thanks{E-mail: ed330@cam.ac.uk}
\\
$^{1}$Astrophysics Group, Cavendish Laboratory, J.J. Thomson Avenue, Cambridge, CB3 0HE, UK\\
$^{2}$Kavli Institue for Cosmology, Madingley Road, Cambridge, CB3 0HA, UK\\
}

\date{Accepted XXX. Received YYY; in original form ZZZ}

\pubyear{2023}

\begin{document}
\label{firstpage}
\pagerange{\pageref{firstpage}--\pageref{lastpage}}
\maketitle

\begin{abstract}
The 21\,cm signal from cosmic hydrogen is one of the most propitious probes of the early Universe.
The detection of this signal would reveal key information about the first stars, the nature of dark matter, and early structure formation.
We explore the impact of an emissive and reflective, or `hot', horizon on the recovery of this signal for global 21\,cm experiments.
It is demonstrated that using physically motivated foreground models to recover the sky-averaged 21\,cm signal one must accurately describe the horizon around the radiometer.
We show that not accounting for the horizon will lead to a signal recovery with residuals an order of magnitude larger than the injected signal, with a log Bayesian evidence of almost 1600 lower than when one does account for the horizon.
It is shown that signal recovery is sensitive to incorrect values of soil temperature and reflection coefficient in describing the horizon, with even a 10\% error in reflectance causing twofold increases in the RMSE of a given fit.
We also show these parameters may be fitted using Bayesian inference to mitigate for these issues without overfitting and mischaracterising a non-detection.
We further demonstrate that signal recovery is sensitive to errors in measurements of the horizon projection onto the sky, but fitting for soil temperature and reflection coefficients with priors that extend beyond physical expectation can resolve these problems.
We show that using an expanded prior range can reliably recover the signal even when the height of the horizon is mismeasured by up to 20\%, decreasing the RMSE from the model that does not perform this fitting by a factor of 9.

\end{abstract}


%%%%%%%%%%%%%%%%%%%%%%%%%%%%%%%%%%%%%%%%%%%%%%%%%%

%%%%%%%%%%%%%%%%% BODY OF PAPER %%%%%%%%%%%%%%%%%%

\section{Introduction}

Creating a timeline of the universe between the period of recombination and the end of reionisation is a necessary step for cosmologists to describe the makeup of the early universe.
Direct observation of cosmic neutral hydrogen remains the most promising tool to understand the universe between \(z \approx 1100\) and \(z \approx 10\).
The hyperfine spin flip of neutral hydrogen produces an emission line at a rest frame of 21\,cm \citep{vandeHulst1945RadiogolvenSpace.}. 
The power of this feature with respect to the radio background as it redshifts through cosmic epochs will provide crucial information about structure formation up to the Epoch of Reionisation.
The depth, position and width of the 21\,cm absorption feature allows us to probe things like early star formation rate \citep[e.g.][]{Schauer2019ConstrainingCosmology}, the initial mass functions of population III stars \citep[e.g.][]{Gessey-Jones2022ImpactSignal}, the nature of X-ray binaries \citep{Das2017High-massAbsorption}, and more exotic physics like dark matter distribution and properties \citep[e.g.][]{Barkana2018StrongSignal}.

We measure the 21\,cm signal using the brightness temperature; that is, the temperature at which a blackbody in thermal equilibrium with an object would have to be to produce the same level of thermal excitation.
The brightness temperature of this signal is several orders of magnitude lower than the radio foreground, making direct observation of the redshifted 21\,cm signal extremely difficult \citep{Shaver1999CanBackground}.
Despite this, a speculative detection of the globally averaged 21\,cm absorption trough was made by the `Experiment to Detect the Global Epoch of Reionization Signal' (EDGES) \citep{Bowman2018AnSpectrum}.
This detection describes the global signal as a flattened Gaussian centred at 76\,MHz with a depth of 0.5\,K.
The flattened nature of the signal, as well as its depth, did not appear to match any existing theory at the time \citep[e.g.][]{Cohen2017ChartingSignal}.
The depth of the signal demanded either an enhanced radio background \citep{Fialkov2019SignatureSpectrum, Mittal2022ImplicationsSurveys}, or a way of cooling the Universe more rapidly than expected due to interactions with dark matter \citep{Liu2019RevivingCosmology}; the flatness possibly being explained by two competing heating mechanisms, such as Lyman-\(\alpha\) photons and cosmic rays becoming dominant at different times \citep{Gessey-Jones2023SignaturesObservables}.

Questions, however, have been raised as to the robustness of the data analysis performed in the experiment \citep{Hills2018ConcernsData}.
Issues may have arisen from nonphysical electron temperatures, a damped sinusoidal systematic \citep{Singh2019TheSpectrum, Sims2019TestingSelection}, and other residuals \citep{Bevins2021Maxsmooth:Cosmology} which may come from beam effects or other distortions such as the ionosphere \citep{Shen2021QuantifyingObservations}.

The `Radio Experiment for the Analysis of Cosmic Hydrogen' (REACH) \citep{deLeraAcedo2022The28}, aims to perform an independent measurement of the sky-averaged 21\,cm signal to either confirm or disprove the EDGES detection \citep{Bowman2018AnSpectrum}.
Utilizing a fully Bayesian data analysis pipeline it aims to model systematics and foregrounds in a more physically motivated manner than has been done previously.

The global 21\,cm absorption trough is predicted to sit between 70 and 200\,MHz \citep[e.g.][]{Pritchard2008EvolutionHistory,Cohen2017ChartingSignal}, which lies in the ranges used by FM radio stations and Digital TV.
This introduces a large amount of possible radio frequency interference (RFI) at a much higher intensity than the 21\,cm signal, which poses a large problem for signal recovery.
This RFI may be mitigated using data analysis tools \citep{Leeney2022AMitigation}, but to further minimise this risk, the radiometer is set up in the Karoo Radio Astronomy Reserve in South Africa, surrounded by mountains on all sides.
While the mountains accomplish the goal of greatly decreasing RFI around the antenna they create a new issue that must be overcome.
It was shown in \citet{Bassett2021LostAnalysis} that the effect of a horizon will be significant on the detection of the 21 cm signal.

For an experiment like EDGES a horizon may be unnecessary to describe, as the polynomials used to fit for their data may have been able to encompass a horizon without describing it specifically.

Any experiment using physically motivated foreground models, for example REACH, however, will require the description of the horizon itself.
This paper focuses on the REACH radiometer \citep{Cumner2021RadioCase} and pipeline \citep{Anstey2020AExperiments}, as this is the collaboration to which the authors belong, but this analysis is applicable to all global 21\,cm experiments using physically motivated foreground models for signal recovery.

Section \ref{sec:methods} deals with the expansion of the REACH pipeline to accommodate for a horizon in data generation and foreground modelling.
Section \ref{sec:results} details the impact of this horizon on signal recovery, the reliance of signal recovery on correct soil parameter estimation, how fitting for soil parameters may liberate one from this reliance, and how this parameter fitting may increase tolerated error in horizon profile mapping. 
In Section \ref{sec:con} we outline our key conclusions and discuss future work to expand the models discussed in this paper.

\section{Methods}
\label{sec:methods}

In this section we describe Bayesian inference (\ref{sec:bayes}), how we simulate a mock data set describing the power given off by a horizon surrounding the REACH antenna (\ref{sec:2.1}), and the methods used to account for this power in the physically motivated foreground models needed to recover the redshifted 21\,cm signal (\ref{sec:2.2}).

\subsection{Bayseian Fitting}
\label{sec:bayes}

The REACH pipeline \citep{Anstey2020AExperiments} relies on Bayesian inference for parameter estimation and model comparison in the recovery of the redshifted 21\,cm signal.
Bayesian inference relies on Bayes' theorem, which states:
\begin{equation}
    P(\theta_\mathcal{M}|\mathcal{D},\mathcal{M}) = \frac{P(\mathcal{D}|\theta_\mathcal{M},\mathcal{M})P(\theta_\mathcal{M}|\mathcal{M})}{P(\mathcal{D}|\mathcal{M})};
\end{equation}
where one can infer the probability of having a set of parameters, \(\theta_\mathcal{M}\), given a set of data, \(\mathcal{D}\), and a proposed model, \(\mathcal{M}\).
This equation may be written more simply to be:
\begin{equation}
    \mathcal{P} = \frac{\mathcal{L}\pi}{\mathcal{Z}},
\end{equation}
where \(\pi\) is the prior, describing our assumptions on the initial probability distribution of the parameters we are estimating, which is updated to the posterior \(\mathcal{P}\).
\(\mathcal{L}\), the likelihood, may be read as the probability of observing the data given a model and a set of parameters using that model.
\(\mathcal{Z}\) is the evidence, representing the probability of observing the data given a model, integrated over all possible parameters that the model could use.

Provided the input data is the same, we can compare the Bayesian evidences of two models to determine which model fits the data with the highest probability, following:

\begin{equation}
    P(\mathcal{M}|\mathcal{D}) = \mathcal{Z}\frac{P(\mathcal{M})}{P(\mathcal{D})},
\end{equation}

\noindent where the ratio of the evidences, weighted by the prior probabilities of the models (which we treat as a uniform distribution), will give the ratio of the probability of each model fitting the data.
The REACH pipeline uses the Nested Sampling \citep{Skilling2006NestedComputationc} algorithm \textsc{Polychord} \citep{Handley2015PolyChord:Sampling, Handley2015Polychord:Cosmology.} to perform this parameter estimation and model comparison.
This algorithm randomly draws a number of parameters from the given prior, for each of which a likelihood is calculated. 
The lowest likelihood points are discarded and the volume of the parameter space shrinks accordingly, updating the priors with a new sample being drawn from the newly constrained prior. 
This iterative shrinking of the prior space is done until a termination criterion has been met and parameters have been determined.
The Bayesian evidences are generated as a byproduct of this and may be further used for model comparison.
For more information on Nested Sampling see \citet{Skilling2006NestedComputationc}. 

\subsection{Data Simulation}
\label{sec:2.1}
\begin{figure}
	\includegraphics[width = \linewidth]{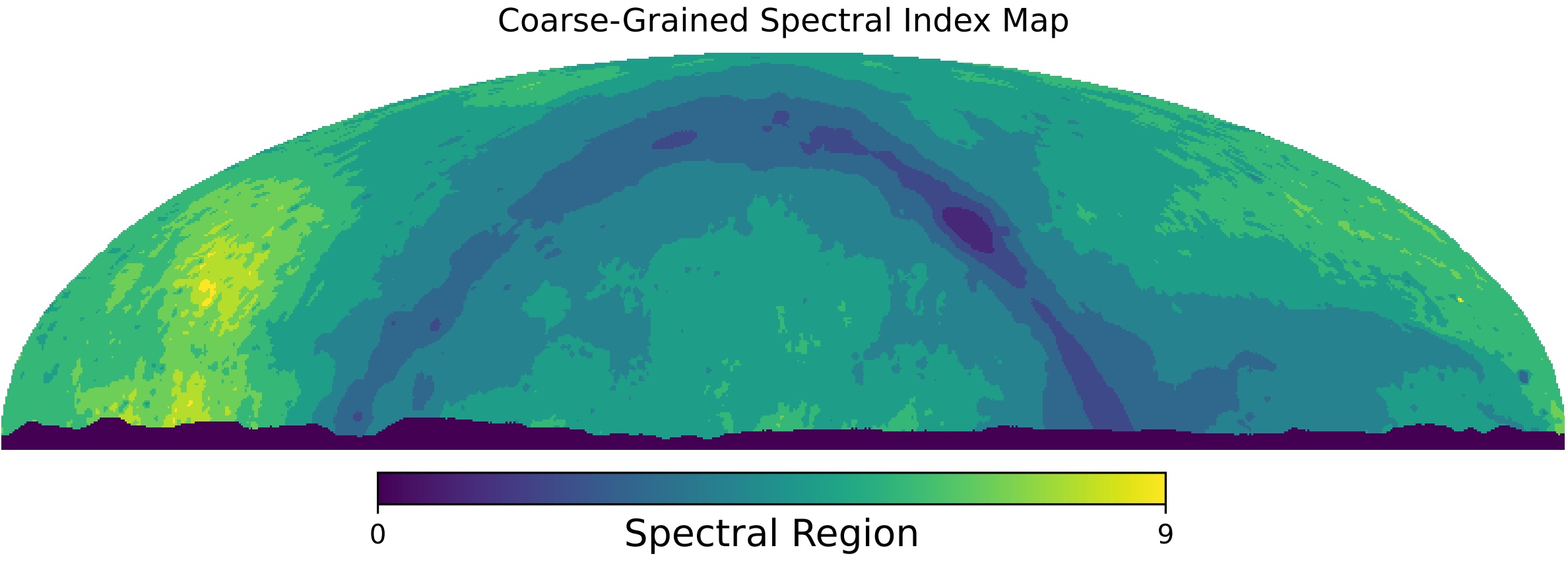}
    \caption{Coarse-grained map of the sky with altitude angle above 0\(^\circ\) in equatorial coordinates. The map is divided into 9 regions where each pixel in a given region is said to have the same spectral index as all other pixels in that region. The horizon here is masked out and assigned spectral region `0'.}
    \label{fig:specregion}
\end{figure}

Using the \textsc{Shapes} algorithm developed by \citet{Bassett2021LostAnalysis} we generate a profile of the horizon around the REACH antenna at \(-30.8387^\circ\)N, \(21.37492^\circ\)E with data from Google Maps based on the Landsat/Copernicus surveys.
We define a horizon mask such that all pixels with an altitude angle \((\theta\)) above that of the horizon for a given azimuthal angle (\(\phi\)) are assigned a value of 0, and any pixel on or under the horizon is assigned a value of 1 (Horizon mask is shown in Figure \ref{fig:specregion}).

Maps of the variation of the spectral indicies across the sky are generated by calculating the spectral index required to map each pixel of the 2008 Global Sky Model (GSM) \citep{deOliveira-Costa2008AGHz} at 408\,MHz onto a corresponding map at 230\,MHz following Equation~\ref{eq:skymap},

\begin{equation}
    \beta(\Omega) = \frac{\text{log}\left(\frac{T_{230}(\Omega) - T_{CMB}}{T_{408}(\Omega)-T_{CMB}}\right)}{\text{log}\left(\frac{230}{408}\right)}.
    \label{eq:skymap}
\end{equation}

\noindent Here we choose a GSM at 230\,MHz to avoid contamination of the cosmological redshifted 21\,cm signal, which at the extremes of theoretical predictions should be extinguished between 200 and 250\,MHz.

We take then the 230\,MHz base map and subtract a flat value of 2.725\,K from each pixel to account for the temperature of the CMB.
This map is then scaled to each frequency according to the previously calculated spectral indicies and rotated according to time and date of a given observation.
Further detail of sky map generation in the REACH pipeline is found in \citet{Anstey2020AExperiments}.

Once these maps have been rotated we mask them by the physical horizon surrounding the radiometer.
In this process we make the assumption that the sky and the horizon sit on the edge of an infinite, flat plane and we may ignore all near-field effects below an altitude angle of 0\(^\circ\).
This assumption is nonphysical, and may cause some issues \citep{Bradley2019AEDGES}, the impacts of near-field soil effects on REACH will be explored in a later work.

Our horizon mask may then be multiplied by some estimate for soil temperature in Kelvin, \(T_\text{soil}\), which will describe simple emission from the soil on the horizon, but will fail to replicate any additional power arising from radio waves reflected from the surface of the soil.

This mask makes the assumption that all light behind the horizon is fully attenuated by the mountains surrounding REACH, and that the temperature of the horizon is constant.
The former is justified as the vegetation surrounding REACH is minimal.
Around REACH, while there is vegetation, there are no large trees in the close foreground that must be relied on to block parts of the sky; thus the attenuation from the mountains can be assumed to be complete.

The latter is a more troublesome approximation, as \(T_\text{soil}\) will decrease or increase at various rates depending on the composition of the horizon and location of that part of the horizon with respect to the sun.
However, for the purposes of this model we believe this is an unnecessary complexity in demonstrating the importance of horizon modelling.

This model also assumes a lack of diffraction.
Light is treated as only moving in straight lines, so any emission from behind the horizon is ignored entirely.
We make this assumption as the ratio of light that is diffracted around the horizon to the amount of light that will be blocked by the horizon is of order \(\lambda/h\), where \(\lambda\) is of order 1m, and \(h\), the height of the mountains around REACH, is of order 1000m.
For much smaller horizons than REACH, diffraction may be a greater challenge, but we address a way of approaching this in Section \ref{sec:con}.

In describing reflection we must make a number of assumptions about the soil, namely its makeup and moisture content, both of which will heavily impact the dielectric permittivity of the soil itself.
As a proxy for the soil in the Karoo reserve we look to examples of better studied soil in other non-sandy desert environments.
The relative dielectric permitivities of soil in the Avra Valley in Arizona were analysed and discussed in \citet{soilperm}.
Using this as an approximation of what we expect to find around the REACH antenna we use a relative dielectric permittivity for Very High Frequency radio waves of order 10.
The reflection coefficient (\(\Gamma\)) of an electromagnetic wave passing from a vacuum into a different medium follows Equation \ref{eq:refco} where we approximate the air to have a relative permittivity (\(\epsilon_r\)) of 1.
We model the reflection as diffuse as it will come from all areas of the sky and the ground, so we can average the angle of incidence to zero, allowing us to model the reflection coefficient as:

\begin{equation}
\label{eq:refco}
    \Gamma = \frac{1- \sqrt{\epsilon_r}}{1 + \sqrt{\epsilon_r}}.
\end{equation}

\noindent Here \(\epsilon_r\) is the relative permittivity of the soil.
Approximating the soil as having an \(\epsilon_r\) of between roughly 5 and 15 depending on soil moisture levels we find a value of \(\Gamma\) to be between -0.4 and -0.6.

Therefore, to account for reflection we take an average of the power per pixel across the entire sky (including blackbody emission from the horizon), add this power onto each pixel of the horizon and multiply it by the magnitude of the reflection coefficient.
This would account for all reflection from the sky and from thermal emission only if the soil were able to see every part of the sky.
This is unphysical.
Thus we also account for the individual parts of the horizon being unable to view the entire sky, i.e. a rock lying on one of the mountains will be unable to see any of the sky that the mountain it is lying on obscures.
We assume the mountains around the telescope to have an incline of 45\(^\circ\), an approximation we make from topographical maps of the area.
We can then multiply the power we have mapped onto the horizon by a factor of 135/180 to account for the regions that remain unseen by that part of the horizon.
This number is specific to the topography of REACH, and the slope of the basin surrounding it.
The value of the incline used has a tolerance of up to \(\pm\)15\(^\circ\) before accurate signal recovery becomes challenging and the depth of the recovered signal is poorly estimated.
While accurate incline estimates are important for accurate signal recovery in our base horizon model the tolerance for precise estimation is loosened greatly once we begin fitting for the reflection coefficient, as in Section \ref{sec:fitting}.

We combine this term, which we deem \(T_{\text{reflection}}\), with our map describing emission to create a snapshot of the sky, which we may integrate over time and solid angle to give a full time-averaged sky map.
This is then convolved with the gain of the beam to give a simulated antenna temperature (\(T_{\text{data}}\)) for an all-sky observation at the REACH site.
It is important to note we assign a value of zero to any part of the antenna beams that have an azimuthal angle below zero.
Reflections of the beam below this angle will be dealt with in a separate paper.

Our data model thus follows: 
\begin{equation}
    \begin{aligned}
        T_{\text{data}}(\nu) = &\;\frac{1}{4\pi}\int_{0}^{4\pi}D(\Omega, \nu) \\ &\times \int_{t_\text{start}}^{t_\text{end}}\left[T_{\text{emission}}(\Omega, \nu, t)+T_{\text{reflection}}(\Omega, \nu, t)\right]d\Omega dt \\ &+ \; \hat{\sigma},
    \end{aligned}
\end{equation}

\noindent where:

\begin{equation}
    \begin{aligned}
        T_{\text{reflection}}(\Omega, \nu) = & \;H(\Omega)\times|\Gamma|\times\left(\frac{135}{180}\right) \\& \times \frac{\int_{0}^{4\pi}T_{\text{emission}}(\Omega, \nu) d\Omega}{\int_{0}^{4\pi}d\Omega},
    \end{aligned}
\end{equation}

\begin{equation}
    \begin{aligned}
        T_{\text{emission}}(\Omega, \nu) =& \;\Big(T_{230}(\Omega)-T_{CMB}(\Omega)\Big)\\&\times\left(\frac{\nu}{230}\right)^{-\beta(\Omega)}
        \Big(1 - H(\Omega)\Big) \\&+  T_{\text{soil}}H(\Omega).
    \end{aligned}
\end{equation}

\noindent \(D(\Omega, \nu)\) is beam directivity at a given frequency, \(H(\Omega)\) is the horizon mask, \(\Omega\) describes solid angle, \(\nu\) is frequency, \(T_\text{soil}\) is the soil temperature in Kelvin, \(T_\text{CMB}\) is the CMB temperature, set to 2.725\,K, and \(\hat{\sigma}\) is experimental noise, which in this case we assume to be a Gaussian white noise set at 25\,mK.

\subsection{Physically Motivated Foreground Model}
\label{sec:2.2}

We follow the framework of \citet{Anstey2020AExperiments} to generate our foreground model.
The sky power is modelled by dividing the sky into \(N\) regions in which we approximate \(\beta\) to be equal across a given region, based on the spectral index map \(\beta(\Omega)\) described in Section \ref{sec:2.1}.
This is shown in Figure \ref{fig:specregion}.
An \(N\) of 1 assumes the entire sky has a constant spectral index, and as \(N\) tends towards number of pixels in the map each pixel will have its own unique spectral index.
This coarse-graining approach allows for greater control of the complexity of the model, with each additional sky region demanding another parameter be fit for in the model.

Maximising the efficiency of the fitting process demands that we calculate the \(N\) sky `chromaticity functions' (\(K_i(\nu)\)) outside of the likelihood.
This means we must approach modelling horizon in the foreground differently to the data generation process.

The coarse-grained spectral index map is rotated according to date and time of observation, and much the same as in the data generation process we mask out the horizon.
Once the horizon has been masked out we map the average power per frequency of each spectral region onto the horizon mask, scaled by the fraction of the sky that it takes up.
Multiplying by \(|\Gamma|\) and amount of the sky that a given part of the horizon can `see' we are given the reflected power per frequency per sky region.
This is a power that we will then scale according to the frequency scaling of the spectral region it arises from.
The sky `chromaticity functions', when multiplied by this sky term, which we denote \(P_\text{sky}\), will account for all power which arises originally from the sky.

\begin{equation}
    \begin{aligned}      
        K_i(\nu) = \; & \frac{1}{4\pi} \int_{0}^{4\pi} D(\Omega, \nu) 
        \times \; M_i(\Omega) \\& \times \; \int_{t_\text{start}}^{t_\text{end}} (T_{230}(\Omega) - T_{\text{CMB}})\,dt d\Omega,
    \end{aligned}
    \label{eq:k}
\end{equation}

\begin{equation}
    \begin{aligned}
       P_{\text{sky}} = \; \left(\frac{\nu}{230}\right)^{-\beta_i}\left(1 + \frac{\frac{1}{4\pi}\int_{0}^{4\pi}H(\Omega)d\Omega\times |\Gamma| \times \left(\frac{135}{180}\right)}{\frac{1}{4\pi}\int_{0}^{4\pi}\left(\sum_{i=1}^{N}M_i(\Omega) + H(\Omega)\right)d\Omega}\right),
    \end{aligned}
    \label{eq:sky}
\end{equation}

\noindent where we follow convention from \citet{Anstey2020AExperiments}. \(\beta_i\) is the spectral index of a given sky region, and \(M_i(\Omega)\) refers to a mask over a given sky region.

We deal with the blackbody emission and self-reflection of the soil separately, as it does not scale with frequency.
We take the horizon mask and multiply it by a given \(T_{\text{soil}}\) to give the blackbody emission term.
To model reflection of the blackbody emission from the soil, we map the emission back onto the horizon mask, accounting for the total fraction of the sky it takes up, and multiplying by \(|\Gamma|\) and the amount of sky that the soil can `see'.
These will account for our blackbody terms, multiplying our blackbody term, \(P_\text{BB}\) by the horizon `chromaticity function' \(J(\nu)\) to give:

\begin{equation}
    \begin{aligned}      
        J(\nu) = \; & \frac{1}{4\pi} \int_{0}^{4\pi} D(\Omega, \nu) 
           \times \; H(\Omega) \\& \times \; \int_{t_{start}}^{t_{end}} (T_{230}(\Omega) - T_{\text{CMB}})\,dt d\Omega,
    \end{aligned}
    \label{eq:j}
\end{equation}

\begin{equation}
    \begin{aligned}
       P_{\text{BB}} =\; T_{\text{soil}}\left(1 + \frac{|\Gamma| \times \left(\frac{135}{180}\right)}{\frac{1}{4\pi}\int_{0}^{4\pi}\left(\sum_{i=1}^{N}M_i(\Omega) + H(\Omega)\right)d\Omega} \right).
    \end{aligned}
    \label{eq:bb}
\end{equation}

These two terms are added together with a flat CMB temperature to give our total model, described by:

\begin{equation}
    \begin{aligned}
        T_{\text{model}} (\nu) = \;  
        \sum_{i=1}^{N} K_i(\nu)P_{\text{sky}}  
        + J(\nu)P_{\text{BB}}  + T_{\text{CMB}}.
    \end{aligned}
\label{eq:model}
\end{equation}

\section{Results}
\label{sec:results}

In this section we discuss the results and implications of a `hot' horizon on the recovery of the 21\,cm signal. 
In subsection \ref{sec:recovery} we discuss the issues that arise when we do not account for, or incorrectly account for the horizon in our foreground models.
In subsection \ref{sec:fixed} we discuss the implications of incorrectly assuming the temperatures and reflection coefficients of the soil on the horizon in correcting for the mismeasurement of the height of a horizon.
subsection \ref{sec:fitting} deals with the utility of allowing the temperature and reflection coefficient of the soil.

\subsection{Recovery of the 21\,cm Signal }
\label{sec:recovery}

In previous works recovering theoretical 21\,cm signals using physically motivated foreground models the horizon is either ignored, or simply treated as something that blocks out sky power without emitting anything itself \citep{Kim2022TheHERA, Hibbard2023FittingCosmology}.

\begin{table}

\centering
\begin{tabular}{llll}
\hline

Parameters & Prior Range\\
\hline
\(F_0\) (MHz) & 50.0  - 200.0\\
Bandwidth (MHz) & 10.0  - 20.0\\
Depth (K) & 0.00 - 0.25\\

\end{tabular}
\caption{Prior ranges for a realistic redshifted Gaussian 21\,cm absorption signal derived from \citet{Cohen2017ChartingSignal}.}
\label{tab:priorrange}
\end{table}

Here we investigate the impact of including a horizon in our mock data set, but failing to properly account for it in our foreground models.
We inject a Gaussian signal with a central frequency of 85\,MHz, bandwidth (here treated as the standard deviation of the Gaussian) of 15\,MHz, and a depth of 0.155\,K into  our mock data set and use our pipeline with 5 distinct models describing the horizon to attempt to recover it.
We detail the prior ranges for a `realistic' redshifted Gaussian 21\,cm absorption feature in Table \ref{tab:priorrange}, derived from \citet{Cohen2017ChartingSignal}.

For each of these models we fit for a Gaussian signal, detailing the central frequency, bandwidth and depth parameters recovered by the model, allowing us to compare these to the true values of the injected signal.
This comparison will give some idea of how accurate the model is.
We use the root mean square error (RMSE) between the injected signal and a Gaussian signal made from the recovered posterior parameters to show the how accurately the recovered signal mirrors the injected one, a lower RMSE indicating a better approximation of the `True' value.
For each of the models we also perform a fit for a non detection, in which the foregrounds are assumed to be the only source of power.
We take the difference of the Log(\(\mathcal{Z}\)) of the fitting our foreground model with an injected Gaussian signal with respect to fitting for just the foregrounds with no signal to give us the \(\delta_{\text{Log}(\mathcal{Z})}\).
This is a measurement of how probable the model believes a detection of a Gaussian signal is when compared to a non-detection.
A \(\delta_{\text{Log}(\mathcal{Z})}\) of 1 would indicate that our model favours a detection with a probability ten times higher than that for a non-detection, with a \(\delta_{\text{Log}(\mathcal{Z})}\) of -1 indicating that the non-detection is favoured by the same amount.
Every increase by 1 in \(\delta_{\text{Log}(\mathcal{Z})}\) corresponds to another order of magnitude by which the detection would be favoured over a non-detection.
Thus, for us to claim that a detection has been made this number cannot be below zero.

We show these results in Table \ref{tab:nohorizoncomp} in which we compare the ability of these 5 different horizon models to recover the 21\,cm absorption trough from Cosmic Dawn.

\begin{figure}
    \centering
    \includegraphics[width = \linewidth]{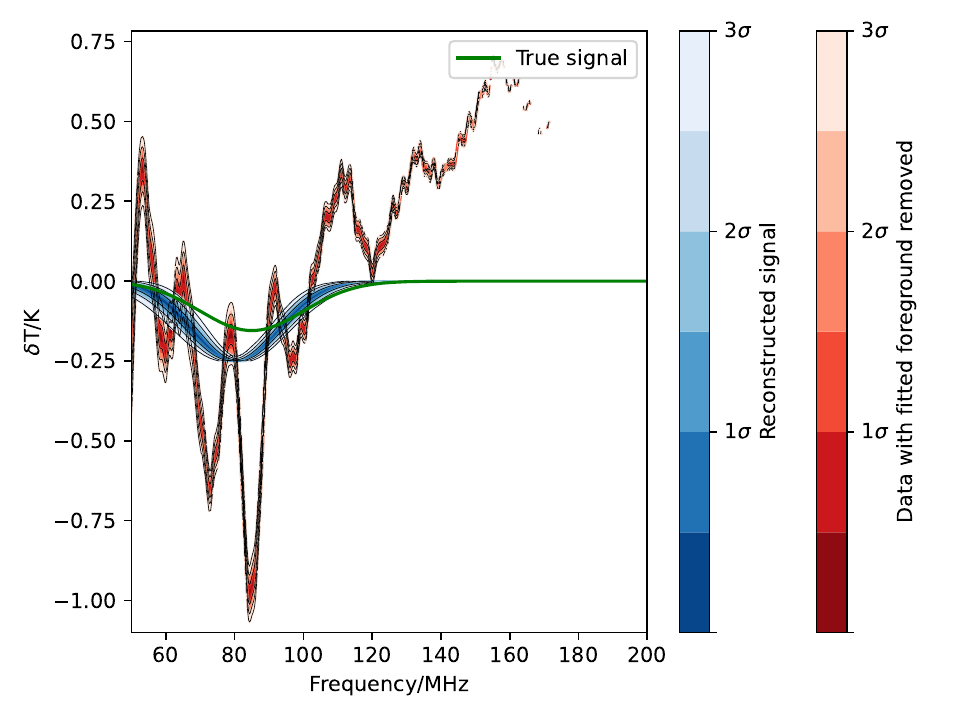}
    \caption{Recovery of a redshifted 21\,cm signal when a horizon is described in the data, but not the foreground models; `No Horizon' model. Injected `True' signal shown in green, with 85\,MHz Central Frequency, 15\,MHz Bandwidth, 0.155\,K Depth. This foreground model fails to accurately recover the injected signal.}
    \label{fig:nohorizon}
\end{figure}

\begin{figure}
    \centering
    \includegraphics[width = \linewidth]{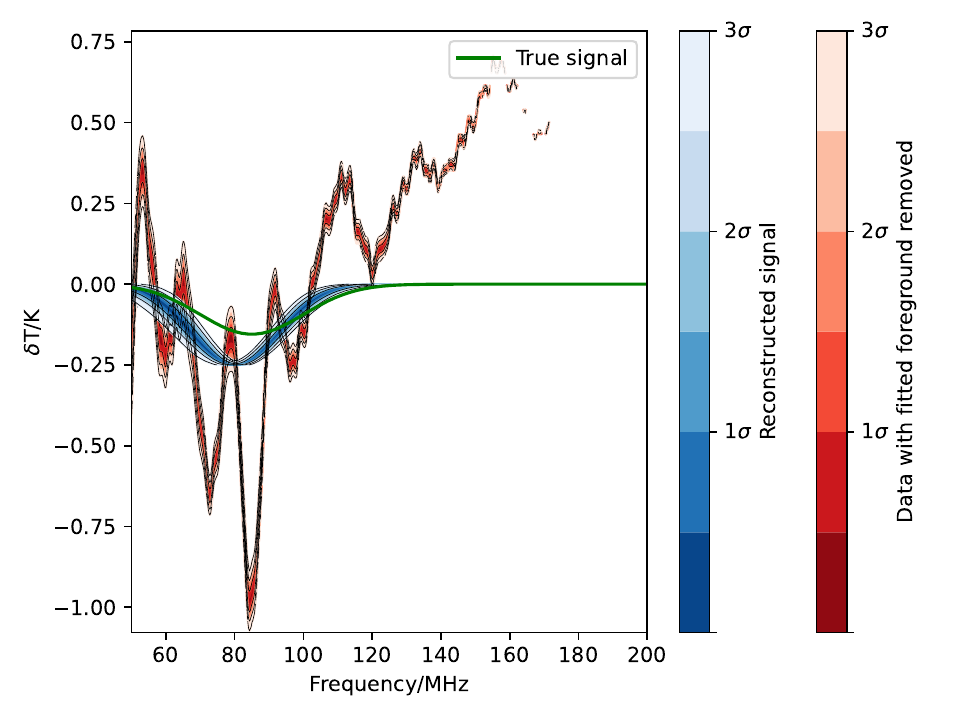}
    \caption{Recovery of a redshifted 21\,cm signal when a horizon is described in the data with foreground models that masks out the horizon from the sky, but does not account for emission or reflection from the soil; `Cold Horizon' model. Injected `True' signal shown in green, with 85\,MHz Central Frequency, 15\,MHz Bandwidth, 0.155\,K Depth.
    This model shows no improvement on the model that does not account for the horizon, once again failing to accurately recover the injected signal.}
    \label{fig:cold}
\end{figure}

\begin{figure}
    \centering
    \includegraphics[width = \linewidth]{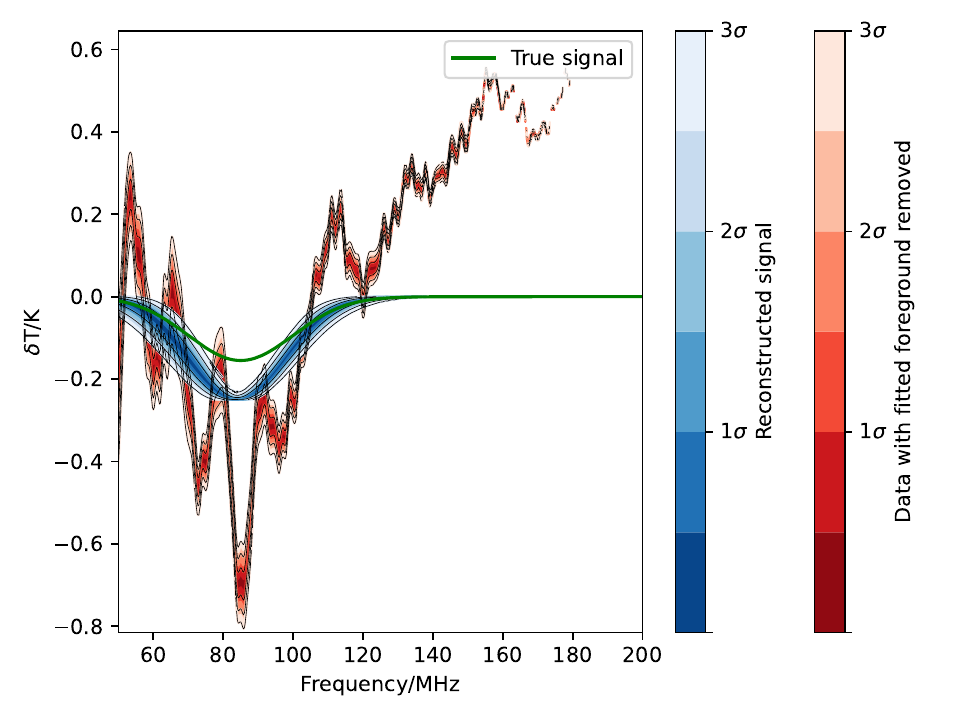}
    \caption{Recovery of a redshifted 21\,cm signal when a horizon is described in the data with foreground models that only account for the reflection of the sky off of the soil; `No Emission' model. Injected `True' signal shown in green, with 85\,MHz Central Frequency, 15\,MHz Bandwidth, 0.155\,K Depth.
    This model shows a decrease in the residuals during signal recovery, but still cannot accurately model the injected signal.}
    \label{fig:noemission}
\end{figure}

\begin{figure}
    \centering
    \includegraphics[width = \linewidth]{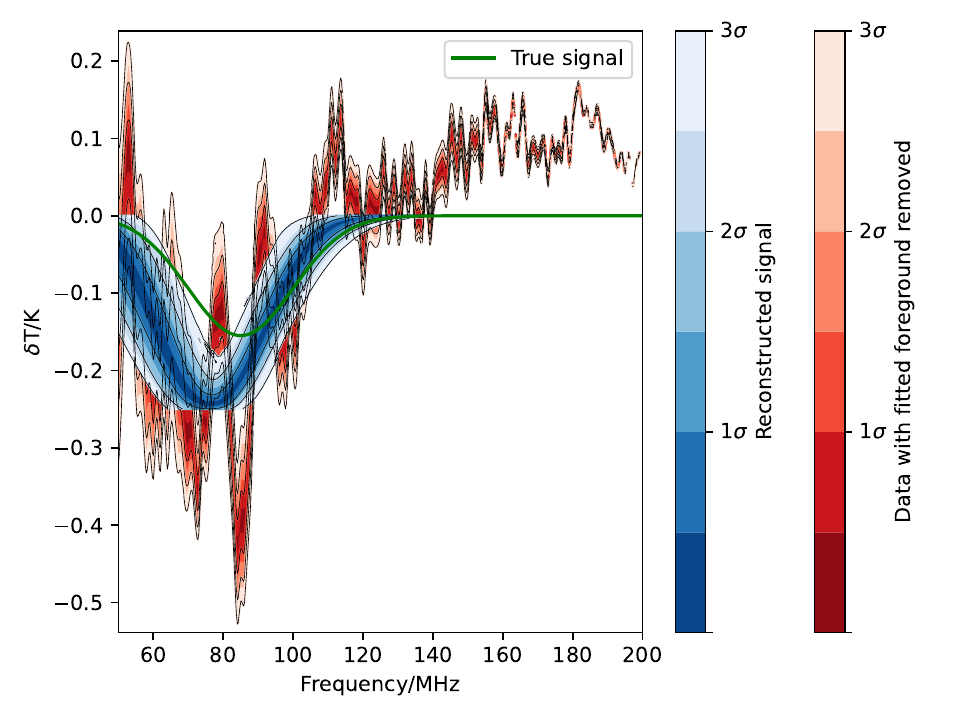}
    \caption{Recovery of a redshifted 21\,cm signal when a horizon is described in the data with foreground models that only account for the blackbody emission from the soil; `No Reflection' model. Injected `True' signal shown in green, with 85\,MHz Central Frequency, 15\,MHz Bandwidth, 0.155\,K Depth.
    This model shows a large step forward in signal recovery, with much smaller residuals than the previous models; it does however still struggle to accurately recover the injected signal. }
    \label{fig:noreflection}
\end{figure}

\begin{figure}
    \centering
    \includegraphics[width = \linewidth]{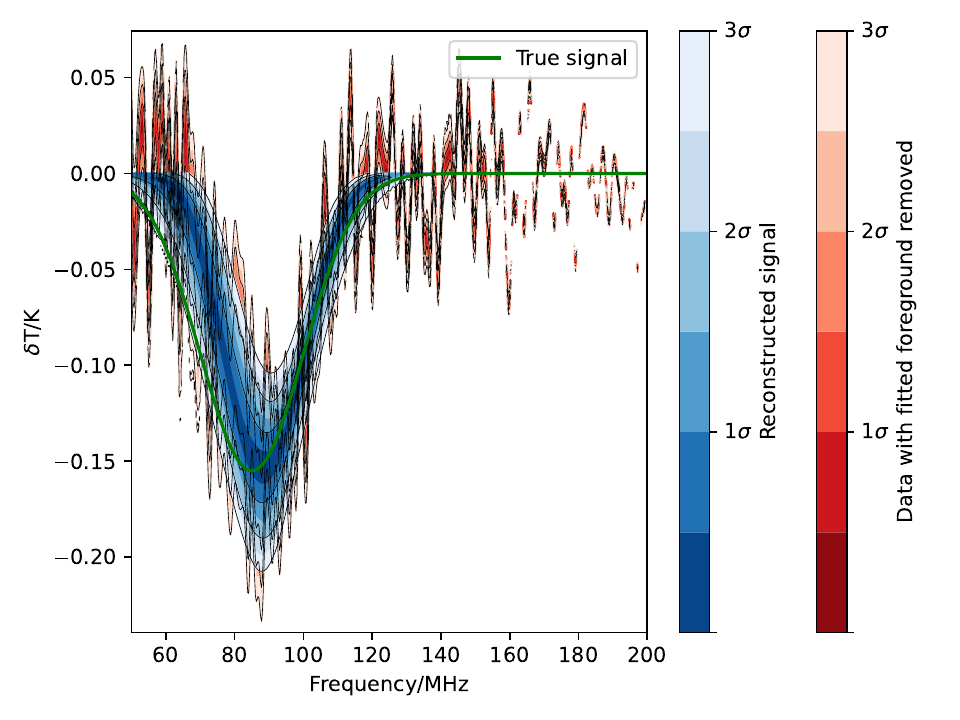}
    \caption{Recovery of a redshifted 21\,cm signal when a horizon is described in the data with foreground models accounting for both blackbody emission from soil and reflected power from the soil and sky; `All' model. Injected `True' signal shown in green, with 85\,MHz Central Frequency, 15\,MHz Bandwidth, 0.155\,K Depth.}
    \label{fig:both}
\end{figure}

\begin{itemize}
    \item{The `No Horizon' model.
    This model fails to account for the horizon entirely, letting \(H(\Omega)\) be a null matrix in Equations  \ref{eq:sky}, \ref{eq:j}, and \ref{eq:bb}.
    As shown in Figure \ref{fig:nohorizon} this saturates our prior for the depth of the signal, and recovers a very biased estimate of the centre frequency with a signal model that has unacceptably large residuals.}
    
    \item{
    The `Cold Horizon' model.
    Here we do include a horizon in our foreground model, but set \(T_{\text{soil}}\) and \(|\Gamma|\) to zero.
    This is an approximation of the approach that previous works have used to describe the horizon \citep{Bassett2021LostAnalysis, Kim2022TheHERA, Hibbard2023FittingCosmology}, treating the horizon as something that attenuates radio waves, instead of being an emitter of any kind.
    We seem to achieve no improvement on the model that ignores the horizon entirely, as shown in Figure \ref{fig:cold}.
    Looking at Table \ref{tab:nohorizoncomp} we see the `Cold Horizon' model appears to struggle to recover the redshifted 21\,cm signal even more than the model that ignores it entirely, with a Log(\(\mathcal{Z}\)) 0.5 lower than the `No Horizon' model and an RMSE 0.0002 higher.
    
    This is not a surprising result.
    The horizon radiates a large amount of power, be it through thermal emission or reflection.
    The signal is buried in the radio foregrounds, so failing to account for the horizon in any way means that the foreground model used will change to account for some emission from the sky in the horizon region.
    This in essence will mimic some of the \(T_\text{reflection}\) term in our data model.
    When one just masks out the horizon and gives it no power we find ourselves even further away from a true description of the foregrounds, and will make signal recovery much more difficult.
    }

    \item{The `No Emission' model.
    Allowing the horizon to reflect the sky, but not providing it with a description of its own thermal emission (T\(_\text{soil}\) = 0, \(|\Gamma|\) = 0.6) does help signal recovery, with a Log(\(\mathcal{Z}\)) increase of almost 600.
    This however, still saturates the depth parameter to the prior limit, seen in Figure \ref{fig:noemission}, due to a large amount of power being unaccounted for in the foreground.
    }

    \item{
    The `No Reflection' model.
    In Figure \ref{fig:noreflection} we account for the thermal emission of the horizon in our foreground models, letting \(T_\text{soil}\) be equal to 300\,K, but not accounting for reflections, keeping \(|\Gamma|\) at 0.
    Here we come much closer to recovering the signal.
    Once again, the signal is biased, and the depth parameter is saturated.
    However, the residuals are greatly reduced. 

    Including thermal emission and ignoring the reflected power gives a much closer approximation to the correct foreground models than when we only consider sky reflection, which, while a large improvement on our `cold' horizon model also entirely saturates the depth priors during signal recovery.
    }

    \item{The `All' model.
    Shown in Figure \ref{fig:both}, when we account for both emission and reflection we find the signal with a Log(\(\mathcal{Z}\)) of 288.3. 
    This is a Log(\(\mathcal{Z}\)) of approximately 1600 more when compared to the model that does not account for the horizon.
    The model that accounts for reflection and emission also has an RMSE of less than half of the no horizon case.
    These values indicate that not only is this model the most favoured in a probabilistic sense, but it is also the model that most accurately recovers our `True' signal. 
    This implies that recovery of the 21\,cm absorption trough using physically motivated foreground models demands realistic horizon modelling.}
\end{itemize}

\begin{table*}
\caption{Comparison of models fitting a redshifted 21\,cm signal when the horizon surrounding the REACH telescope is left unaccounted for, has parts of its power unaccounted for, or is accounted for correctly.
The No Horizon model describes a foreground model which does not account for the Horizon.
The `Cold' Horizon model masks the horizon out of the sky model, but sets \(T_\text{soil}\) and \(|\Gamma|\) to be 0.
The No Emission model describes a model in which the horizon is able to reflect sky power, but does not thermally emit any power, setting \(T_\text{soil}\) as 0\,K and \(|\Gamma|\) as 0.6.
The No Reflection model describes a model that does not reflect any sky or horizon power, but is able to thermally emit power, with \(T_\text{soil}\) at 300\,K and \(|\Gamma|\) set to 0.
The `All' model describes the horizon both emitting power thermally, and reflecting both sky power and the power of the horizon itself; here \(T_\text{soil}\) is at 300\,K and \(|\Gamma|\) is 0.
The inserted mock signal has an 85\,MHz Central Frequency, 15\,MHz Bandwidth, 0.155\,K Depth. 
In the mock data soil temperature is set to 300\,K and the magnitude of the reflection coefficient is set to 0.6. 
T and \(|\Gamma|\) refer to the soil temperature and magnitude of the reflection coefficients respectively in our foreground model.
\(\mathcal{Z}_\text{Gaussian}\) is the Bayesian evidence of trying to fit the injected signal to a Gaussian, and \(Z_\text{No 21}\) is the Bayesian evidence when we try to model for our data having no 21\,cm signal.
\(\delta_{\text{Log}(\mathcal{Z}})\) is the difference in evidence between these models.
RMSE refers to the root mean squared error when comparing the injected mock signal to one that we generate using the posterior averages that our Gaussian model suggests.}
\label{tab:nohorizoncomp}
\centering
\begin{tabular}{lllllllllll}
\hline
& T & \(|\Gamma|\) & F\(_0\) (MHz) & Bandwidth (MHz) & Depth (K) & Log(\(\mathcal{Z}_\text{Gauss}\)) & Log(\(\mathcal{Z}_\text{No 21}\)) & \(\delta_{\text{Log}(\mathcal{Z})}\) & RMSE\\
\hline
\hline
\hline
Mock Signal & 300 & 0.6 & 85.0 & 15.0 & 0.155\\
\hline
Models\\
\hline
No Horizon & N/A & N/A & 80.3\(\pm{1.1}\) & 12.7\(\pm{0.9}\) & 0.248\(\pm{0.002}\) & -1194.7\(\pm{0.4}\)  & -1326.6\(\pm0.4\) & 131.9\(\pm0.6\) & 0.0372\\
Cold Horizon & 0 & 0 & 80.2 \(\pm1.1\) & 12.7\(\pm1.0\) & 0.248\(\pm0.002\) & -1196.4\(\pm0.4\) & -1327.8\(\pm0.4\) & 131.4\(\pm0.6\) & 0.0374
\\
No Emission & 0 & 0.6 & 83.9\(\pm1.3\) & 13.7\(\pm1.2\) & 0.247\(\pm0.003\) & -667.3\(\pm0.4\) & -782.5\(\pm0.4\) & 115.2\(\pm0.6\) & 0.0348
\\No Reflection & 300 & 0 & 76.5\(\pm{2.7}\)& 15.1\(\pm{1.9}\) & 0.240\(\pm0.010\)  & 114.0\(\pm{0.3}\) & 93.2\(\pm0.4\) & 20.8\(\pm0.6\) & 0.0477
\\All & 300 & 0.6 & 88.9\(\pm{1.3}\) & 12.1\(\pm{1.1}\) & 0.154\(\pm{0.016}\) & 288.3\(\pm{0.4}\) & 242.8\(\pm0.4\) & 45.5\(\pm0.6\) & 0.0169\\
\hline
\end{tabular}
\end{table*}


\subsection{Investigation into Effects of Temperature and Reflection Coefficient}
\label{sec:fixed}

Until this point we have assumed we are able to perfectly predict both the temperature and reflection coefficient of the soil on the REACH horizon.
Practically this is difficult.

The temperature of the soil will vary with time, and while it is possible to set up temperature probes around the horizon of the antenna, this is difficult both fiscally and in terms of the human resources it would require.
The Karoo is a radio quiet reserve, so temperature probes cannot be remote, and must be collected from around the reserve manually upon each observation.
\(|\Gamma|\) is even more difficult to determine. 
This is heavily dependent on moisture levels of the soil, how deep the moisture penetrates the soil, and specific soil makeup across the mountains.
This will all vary with weather and location around the mountain, which makes precise estimates of \(|\Gamma|\) year round unfeasible.

Using the same model throughout for temperature and reflection coefficient in our model of data generation (300\,K and 0.6 respectively), we examine the impact of incorrectly predicting the soil temperature and reflection coefficient in our foreground models.

%\begin{figure}
%    \centering
%    \includegraphics[width=\linewidth]{Images/test.pdf}
%    \caption{Comparison of log evidences of foreground models with varying soil temperatures and reflection coefficients recovering a redshifted 21\,cm signal using the REACH pipeline and a log spiral antenna. In the data model the \(T_\text{soil}\) and \(|\Gamma|\) are set to 300\,K and 0.6 respectively}
%    \label{fig:temprefcomp}
%\end{figure}

%\begin{figure}
%    \centering
%    \includegraphics[width=\linewidth]{Images/test2.pdf}
%    \caption{Comparison of RMSE of foreground models with varying soil temperatures and reflection coefficients recovering a redshifted 21\,cm signal using the REACH pipeline and a log spiral antenna. In the data model the \(T_\text{soil}\) and \(|\Gamma|\) are set to 300\,K and 0.6 respectively}
%    \label{fig:temprefcomprmse}
%\end{figure}

\begin{figure*}
\captionsetup[subfigure]{justification=centering, skip=-30pt}
\begin{subfigure}{.49\textwidth}
  \centering
  \captionsetup{width=\linewidth}
  % include first image
  \includegraphics[width=\textwidth]{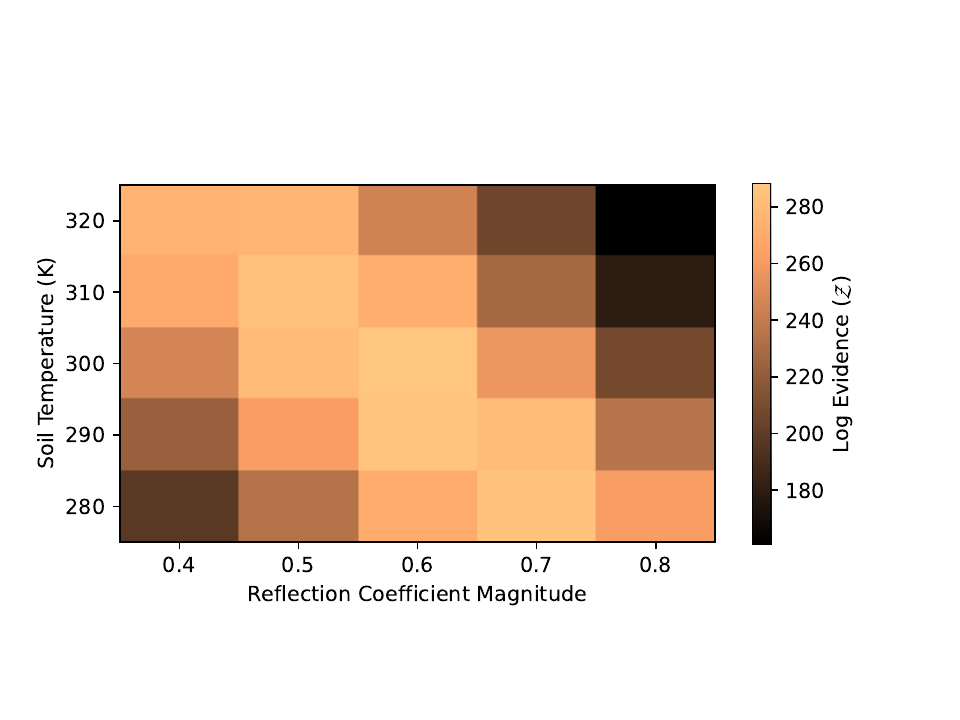}  
  \label{fig:logev}
  \subcaption{}
\end{subfigure}
\begin{subfigure}{.49\textwidth}
  \centering
  \captionsetup{width=\linewidth}
  % include first image
  \includegraphics[width=\textwidth]{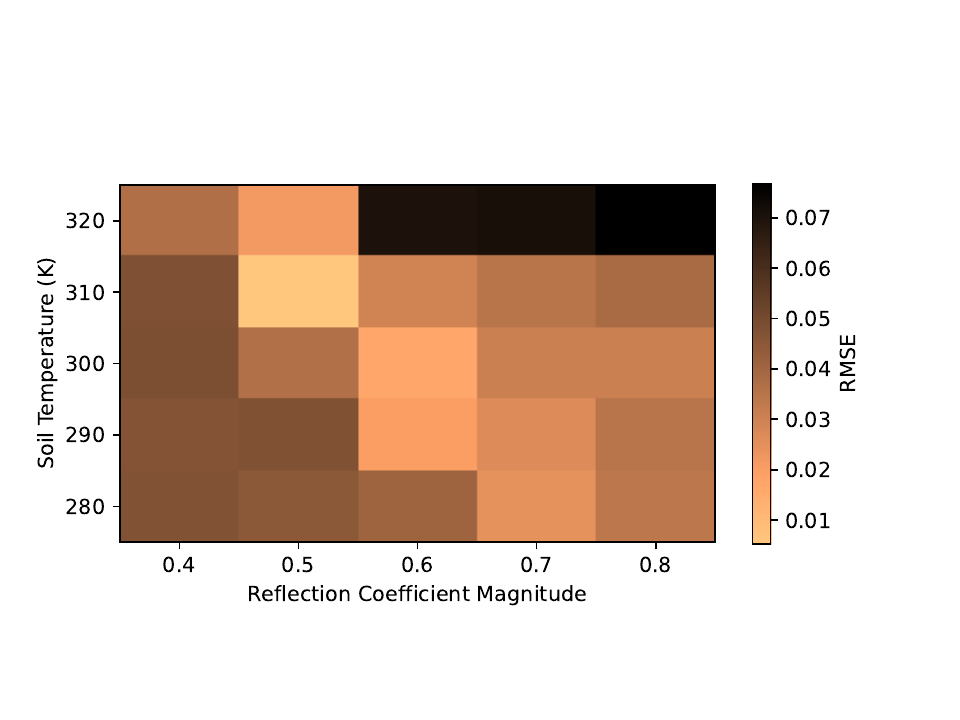}  
  \label{fig:rmsecomp}
  \subcaption{}
\end{subfigure}
\caption{Comparison of foreground models with varying soil temperatures and reflection coefficients recovering a redshifted 21\,cm signal using the REACH pipeline and a log spiral antenna. In the data model the \(T_\text{soil}\) and \(|\Gamma|\) are set to 300\,K and 0.6 respectively. (a) shows the log Bayesian evidence as a function of changing foreground parameters, and (b) shows the corresponding change in root mean squared error of the recovered signal compared to the `True' signal.}
\label{fig:temprefcomp}
\end{figure*}

It may be observed from Figure \ref{fig:temprefcomp} (for full details see Table \ref{tab:changingtempandrefco}) that our misjudging the values of \(T_\text{soil}\) and \(|\Gamma|\) in the foreground models will cause problems for signal recovery.
It does, however, demonstrate the expected link between values chosen for \(T_\text{soil}\) and \(|\Gamma|\).
Both will describe the amount of power emitted from the horizon; increasing \(T_\text{soil}\), leading to  overestimation of emission power from the horizon may be compensated for by decreasing the value of \(|\Gamma|\) and vice-versa.

This is not a perfect fix.
\(|\Gamma|\) is contained in Equation \ref{eq:sky}, while \(T_\text{soil}\) is not. 
This is an issue as this sky term in our model is the only one that gets scaled by frequency, so neither one term can entirely correct for the other.
As a result we will need very accurate estimates of both \(T_{\text{soil}}\) and \(|\Gamma|\) if we are correctly recover the 21\,cm signal.

\subsection{Soil Temperature and Reflection Coefficient Fitting}
\label{sec:fitting}

As shown in Section \ref{sec:fixed}, by fixing values of \(T_{\text{soil}}\) and \(|\Gamma|\) as reasonable estimates of the correct parameters\footnote{With \(T_{\text{soil}}\) being within 20\,K, and \(|\Gamma|\) being within 0.2 of the values we use in our data model} which are slightly mismatched from the mock data we may yield a non-detection, or detection of a signal with very different parameters to the true one.

We can potentially mitigate for these issues by fitting for both \(T_\text{soil}\) and \(|\Gamma|\) as additional parameters in our model.
Setting the priors of our model to be uniform for \(T_\text{soil}\) and \(|\Gamma|\) between 275 and 325\,K, and for between 0.4 and 0.7 respectively, we explore the ability of this model to recover a range of signals.

\begin{figure}
    \centering
    \includegraphics[width = \linewidth]{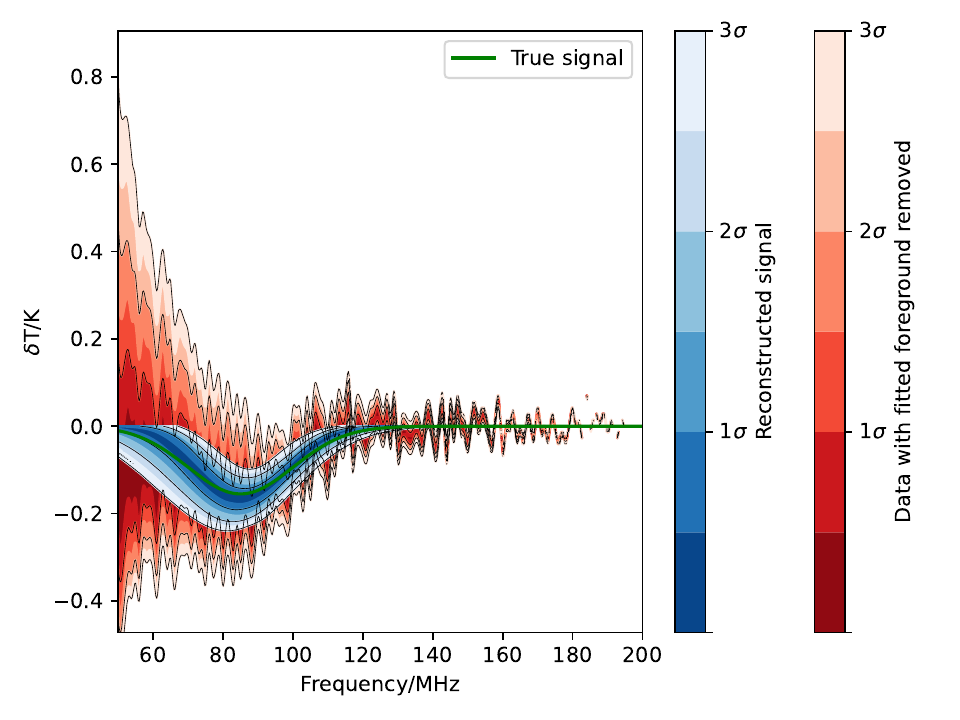}
    \caption{Recovery of a theoretical redshifted Gaussian 21\,cm signal using a log spiral antenna and the REACH pipeline in which we fit for soil temperature and reflection coefficient as additional parameters. Injected `True' signal is shown in green, with 85\,MHz Central Frequency, 15\,MHz Bandwidth, 0.155\,K Depth.}
    \label{fig:guassallfit}
\end{figure}

\begin{figure}
    \centering
    \includegraphics[width = \linewidth]{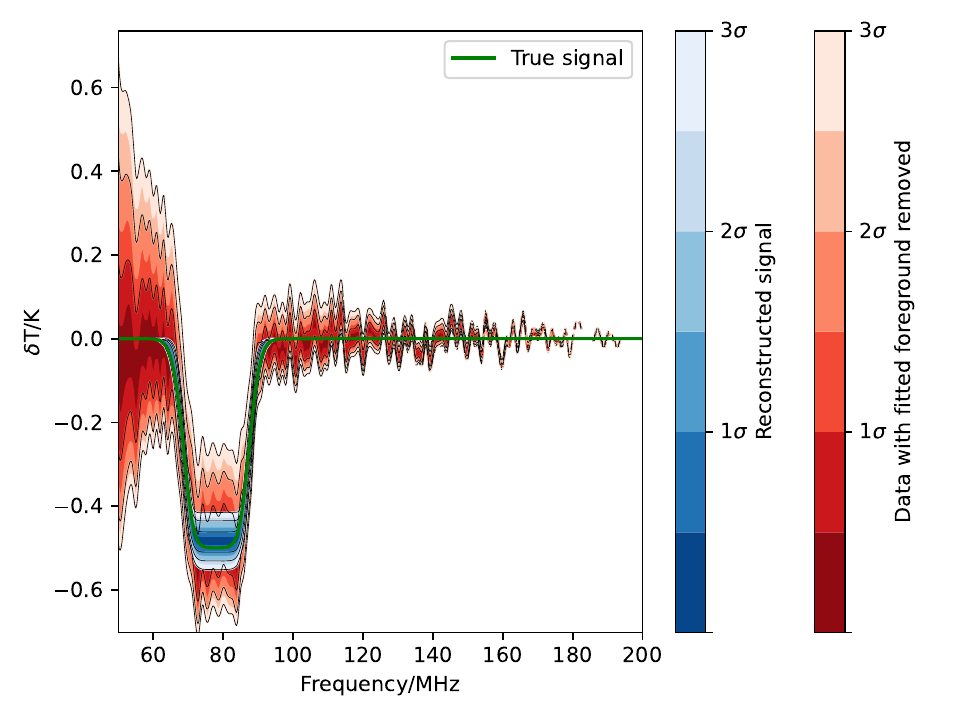}
    \caption{Recovery of a theoretical redshifted flattened Gaussian 21\,cm signal using a log spiral antenna and the REACH pipeline in which we fit for soil temperature and reflection coefficient as additional parameters. Injected `True' signal is the recovered signal from EDGES, shown in green, with 76\,MHz Central Frequency, 19\,MHz Bandwidth, 0.5\,K Depth, and a Flattening Factor of 7.}
    \label{fig:fgaussallfit}
\end{figure}

The model is able to accurately recover Gaussian and flattened Gaussian signals, as in Figures \ref{fig:guassallfit} and \ref{fig:fgaussallfit}, with a low RMSE, but much higher residuals at lower frequencies than we saw for models in which we fixed \(T_\text{soil}\) and \(|\Gamma|\) to specific, correct, values.

When we fit for the reflection coefficient, however, we see very large residuals at low frequencies.
These residuals arise in Equation \ref{eq:sky} where \(|\Gamma|\) is multiplied by \(\nu^{-\beta_i}\), preferentially increasing the residuals for lower frequencies.
While these residuals are large they are non-degenerate with the signal we recover, so do not create any issues for the model itself.

\begin{table*}
\footnotesize
\caption{An examination of how the pipeline copes with fitting soil temperature and reflection coefficient parameters with a range of varying signal parameters and soil properties generated using a pseudo random number generator. `Fixed Parameters' refers to \(T_\text{soil}\) and \(|\Gamma|\) being hard coded into the foreground model to have the same values as the data model, `Fitted' allows these parameters to be additional parameters that we fit for in the foreground model. \(\mathcal{Z}_\text{Gaussian}\) is the Bayesian evidence of trying to fit the injected signal to a Gaussian, and \(\mathcal{Z}_\text{No 21}\) is the Bayesian evidence when we try to model for our data having no 21\,cm signal.
\(\delta_{\text{Log}(\mathcal{Z}})\) is the difference in evidence between these models.
RMSE refers to the root mean squared error when comparing the injected mock signal to one that we generate using the posterior averages that our Gaussian model suggests.}
\label{tab:fittesting}
\centering
\begin{tabular}{llllllllll}
\hline
& F\(_0\) (MHz) & Width (MHz) & Depth (K) & T(K) & \(|\Gamma|\) & Log\((\mathcal{Z}_{\text{Gauss}}\)) & Log\((\mathcal{Z}_{\text{No 21}}\)) & \(\delta_{\text{Log}(\mathcal{Z})}\) & RMSE\\
\hline
\hline
\hline
True & 85.0 & 15.0 & 0.155 & 300.0 & 0.60 \\
Fixed Parameters & 88.9\(\pm1.3\) & 12.1\(\pm1.1\) & 0.154\(\pm0.016\) &  &  & 288.3\(\pm0.4\) & 244.4\(\pm0.4\) & 43.9\(\pm0.6\) & 0.0169\\
Fitted & 87.9\(\pm2.0\) & 13.1\(\pm1.5\) & 0.174\(\pm0.021\) & 294.9\(\pm5.3\) & 0.61\(\pm0.04\) & 287.0\(\pm0.4\) & 253.7\(\pm0.4\) & 33.3\(\pm0.6\) & 0.0120\\
\hline
True & 180.0 & 10.7 & 0.233 & 316.0 & 0.54\\
Fixed Parameters & 180.4\(\pm0.4\) & 11.4\(\pm0.5\) & 0.232\(\pm0.007\) &  &  & 282.8\(\pm0.4\) & 113.5\(\pm0.4\) & 169.3\(\pm0.6\) & 0.0044\\
Fitted & 180.5\(\pm0.4\) & 11.4\(\pm0.5\) & 0.230\(\pm0.009\) & 321.1\(\pm5.5\) & 0.49\(\pm0.02\) & 279.9\(\pm0.4\) & 169.5\(\pm0.4\) & 110.4\(\pm0.6\) & 0.0043\\
\hline
True & 53.0 & 12.1 & 0.057 & 293.0 & 0.41 \\
Fixed Parameters & 138.8\(\pm47.9\) & 15.0.2\(\pm3.0\) & 0.014\(\pm0.012\) &  &  & 289.9\(\pm0.4\) & 293.8\(\pm0.4\) & -4.9\(\pm0.6\) & 0.0185\\
Fitted & 101.0\(\pm41.1\) & 14.2\(\pm2.8\) & 0.022\(\pm0.015\) & 291.4\(\pm4.4\) & 0.39\(\pm0.03\) & 286.7\(\pm0.4\) & 291.6\(\pm0.4\) & -4.9\(\pm0.6\) & 0.0193\\
\hline
True & 141.0 & 17.2 & 0.152 & 277.0 & 0.68\\
Fixed Parameters & 141.1\(\pm0.8\) & 17.4\(\pm1.0\) & 0.156\(\pm0.006\) &  &  & 283.5\(\pm0.4\) & 149.1\(\pm0.4\) & 134.4\(\pm0.6\) & 0.0023\\
Fitted & 140.8\(\pm0.8\) & 15.5\(\pm1.1\) & 0.151\(\pm0.007\) & 276.5\(\pm0.4\) & 0.63\(\pm0.02\) & 276.8\(\pm0.4\) & 183.7\(\pm0.4\) & 93.1\(\pm0.6\) & 0.0063\\
\hline
True & 126.0 & 13.8 & 0.069 & 322.0 & 0.48\\
Fixed Parameters & 126.1\(\pm2.3\) & 16.3\(\pm2.7\) & 0.072\(\pm0.007\) &  &  & 283.7\(\pm0.4\) & 245.1\(\pm0.4\) & 38.6\(\pm0.6\) & 0.0051\\
Fitted & 128.6\(\pm1.8\) & 12.2\(\pm1.9\) & 0.067\(\pm0.007\) & 325.2\(\pm4.0\) & 0.43\(\pm0.02\) & 282.9\(\pm0.4\) & 253.6\(\pm1.1\) & 29.3\(\pm1.2\) & 0.0050\\
\hline
True & 72.0 &  18.0 &  0.211 & 285.0 & 0.59 \\
Fixed Parameters & 79.5\(\pm2.0\) & 14.6\(\pm1.7\) & 0.162\(\pm0.020\) &  &  & 289.3\(\pm0.4\) & 266.1\(\pm0.4\) & 13.2\(\pm0.4\) & 0.0379\\
Fitted & 81.3\(\pm2.8\) & 14.5\(\pm1.9\) & 0.175\(\pm0.023\) & 276.7\(\pm5.3\) & 0.63\(\pm0.4\) & 286.9\(\pm0.4\) & 263.4\(\pm0.4\) & 23.5\(\pm0.6\) & 0.0397\\
\hline
\hline

True & 0.0 &  0.0 &  0.000 & 0.0 & 0.00 \\
Fixed Parameters & 121.0\(\pm41.0\) & 14.6\(\pm3.0\) & 0.016\(\pm0.012\) &  &  & 288.3\(\pm0.8\) & 289.9\(\pm0.4\) & -1.6\(\pm0.9\) & 0.0066\\
Fitted & 117.1\(\pm47.6\) & 14.8\(\pm2.8\) & 0.022\(\pm0.017\) & 301\(\pm5.3\) & 0.56\(\pm0.03\) & 285.8\(\pm0.4\) & 288.8\(\pm0.4\) & -3.0\(\pm0.6\) & 0.0092\\

\hline
\label{tab:fittedvsperfect}
\end{tabular}
\end{table*}

In Table \ref{tab:fittedvsperfect} we compare the ability of our new model, fitting for parameters, to a `perfect' model, where \(T_\text{soil}\) and \(|\Gamma|\) are fixed to match the input values in our mock data\footnote{It is important to note for this comparison we can only make comparisons between models of a given signal, the evidences relating from one signal to another will be incomparable as the mock data we are fitting to will be different.}.
To test this we generate 6 Gaussian signals with pseudo-random parameters designed to stretch the model beyond the simple 85\,MHz, 15\,MHz, 0.155\,K signal we have been using up to this point.

This model in which we fit for \(T_\text{soil}\) and \(|\Gamma|\) performs consistently as well as the model in which we have fixed parameters.
The RMSE values of the two models yield similar results, and the \(\delta_{\text{Log}(\mathcal{Z})}\) of the two models indicate the same ability of the models to recover a signal.

In the cases where the fixed parameter version of our model finds a signal, the fitted one will too.
However, when one fails the other does the same.
A signal that cannot be found with the fixed parameter model will not be found with the fitted model.
This is most notable when the we attempt to recover signals centred at 53 and 72\,MHz, falling at the very low end of the REACH observation band where the Gaussian signal does not fit entirely within the band.
The 53\,MHz signal is not consistently detected, the pipeline giving a higher evidence for a non detection.
The 72\,MHz signal is found, but the pipeline does not recover the parameters to within reasonable error in either the fitted or hard coded cases.
This detection has a very high RMSE, meaning we must be very careful in assuming the validity of any signals recovered around this range.

The pipeline being unable to properly recover signals that do not entirely fall within the observing band of REACH is not unexpected.
A signal that does not sit fully within the band will be more degenerate with the foregrounds and will be fitted poorly.

Once the signal sits more comfortably in the observation band both the traditional fixed model, and the model were we fit for \(T_\text{soil}\) and \(|\Gamma|\) accurately recover the signal.
Crucially, when we fit for a model that assumes no signal we still see a much greater evidence for the model that contains a signal, which is to say that \(T_\text{soil}\) and \(|\Gamma|\) being fitted will not arbitrarily increase the evidence of a fit to where it becomes impossible to determine detection from non-detection.

To confirm this we also perform a test in which we inject no signal, and fit this to a Gaussian.
It can be seen that even when we are fitting the additional parameters of \(T_\text{soil}\) and \(|\Gamma|\) we do not artificially recover a Gaussian detection, with the Bayesian evidence correctly favouring a non-detection.

\subsection{Soil Temperature and Reflection Coefficient Fitting as a Way of Correcting for Horizon Measurement Errors}

We have allowed ourselves to deal with any error in the measurement of \(T_\text{soil}\) and \(|\Gamma|\) using the fitting process defined in Section \ref{sec:fitting}.
However, all models to this point assume that the physical height of the horizon and its projection onto the sky is measured without error.
This is not a reasonable assumption.

The model being able to fit for both \(T_\text{soil}\) and \(|\Gamma|\) may be able to compensate for errors in the measurement of the horizon.
One might assume that artificially increasing \(T_\text{soil}\) would cause the model to perceive the horizon to be higher than it actually is, or by increasing \(|\Gamma|\) the model will see more of the sky than it really does.
We examine this na\"ive assumption in this section.

To explore the functionality of \(T_\text{soil}\) and \(|\Gamma|\) fitting as a way to mitigate error in horizon height we expand the prior range beyond physical expectation.
We allow \(T_\text{soil}\) to sit between 200 and 400\,K.
We allow \(|\Gamma|\) to have any value between 0.3 and 0.9.
We will then systematically create an artificial error in the horizon we use for our foreground modelling.
We multiply the altitude angle, \(\theta\), in the horizon mask of the foreground model, \(H(\Omega)\), by some scaling factor, \(S\), to increase or decrease the height of the horizon in the foreground model with respect to the mock data.

\begin{figure}
    \centering
    \includegraphics[width =\linewidth]{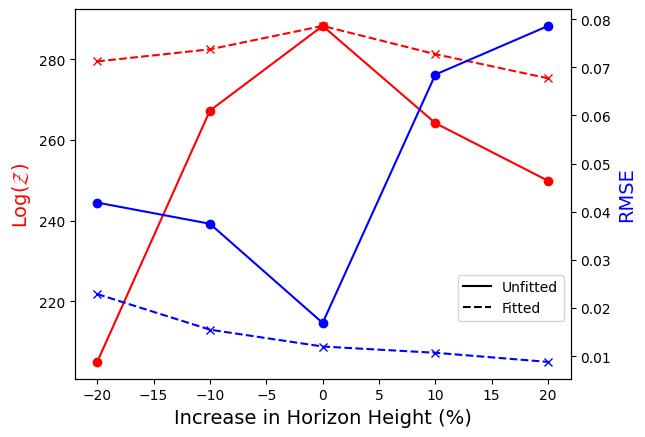}
    \caption{An exploration into how fitting for temperature and reflection coefficient of soil surrounding the REACH antenna allows for the incorrect measurement of height of the horizon in the foreground models. We inject a signal with an 85M\,Hz Central Frequency, 15\,MHz Width, 0.155\,K Depth and our data model surrounds the telescope with soil at 300\,K with a \(|\Gamma|\) of 0.6. Log(\(\mathcal{Z}\)) is the Bayesian evidence of the model fitting the injected signal to a Gaussian. RMSE refers to the root mean squared error when comparing the injected mock signal to one that we generate using the posterior averages that our Gaussian model suggests. The `Unfitted' model fixes values for \(T_\text{soil}\) and \(|\Gamma|\) to be 300\,K and 0.6 respectively. The `Fitted' model fits for \(T_\text{soil}\) and \(|\Gamma|\) as additional parameters with priors between 200-400\,K and 0.3-0.9 respectively.}
    \label{fig:fittingtocorrect}
\end{figure}

We examine the results of this in Figure \ref{fig:fittingtocorrect} (for full details see Table \ref{tab:fittingtocorrect}), where we compare how the pipeline deals with incorrect horizon height estimates when we input the same values of \(T_\text{soil}\) and \(|\Gamma|\) as was used to create our mock data set versus when we allow for those parameters to be fitted with an increased prior range.
Here we try to recover the 85\,MHz, 15\,MHz, 0.155\,K signal at 300\,K with a \(|\Gamma|\) of 0.6.
We set \(S\) to be 0.8, 0.9, 1, 1.1 and 1.2 to give a deviation in the horizon height measurement in the foreground models of up to 20\% from the mock data.

\begin{figure*}
\captionsetup[subfigure]{justification=centering, skip=-10pt}
\begin{subfigure}{.49\textwidth}
  \centering
  \captionsetup{width=.98\linewidth}
  % include first image
  \includegraphics[width=\textwidth]{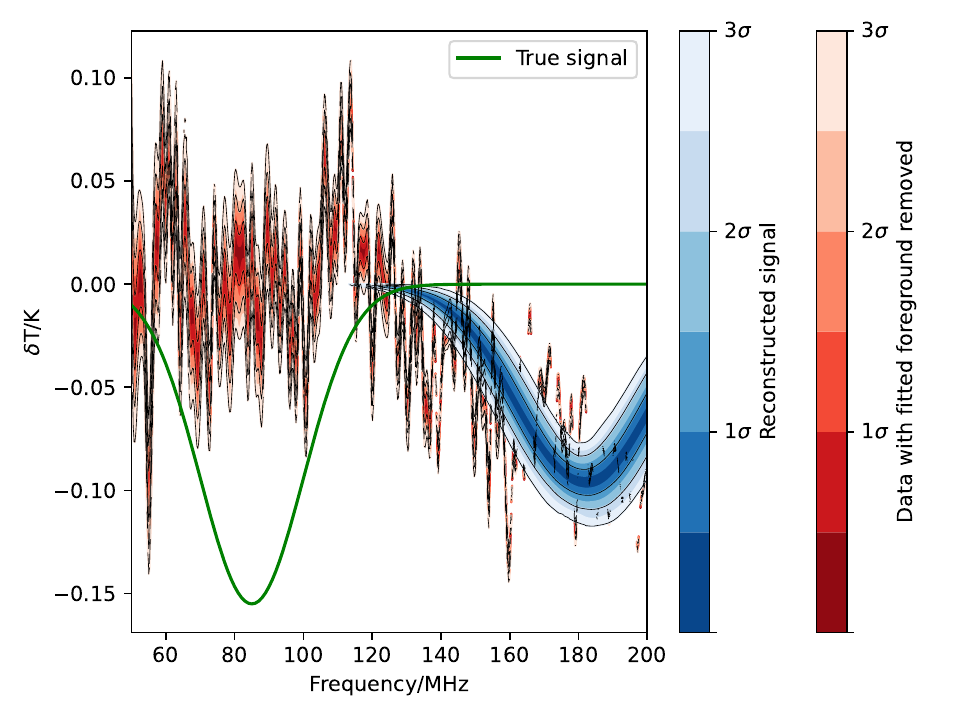}  
  \label{fig:20perno}
  \subcaption{}
\end{subfigure}
\begin{subfigure}{.5\textwidth}
  \centering
  \captionsetup{width=.98\linewidth}
  % include first image
  \includegraphics[width=\textwidth]{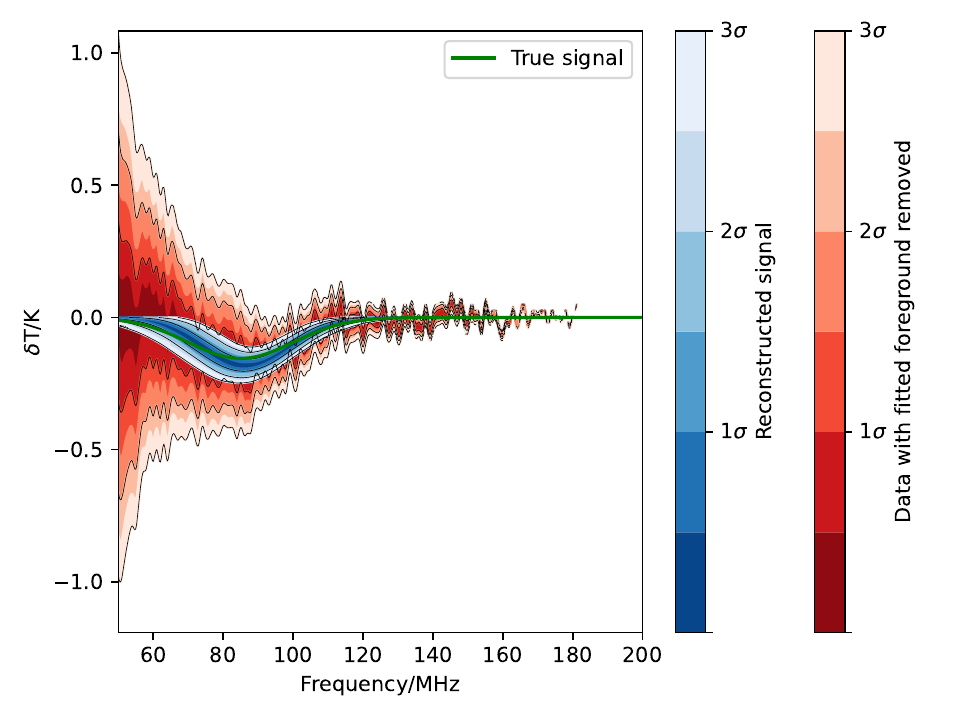}  
  \label{fig:20peryes}
  \subcaption{.}
\end{subfigure}
\caption{A comparison of recovery of a theoretical redshifted 21\,cm signal (F\(_0\) = 85\,MHz, Bandwidth = 15\,MHz, Depth = 0.155\,K) using the REACH pipeline with a log spiral antenna and a foreground model that assumes the horizon to be 20\% higher than the model used in data generation.
`True' signal is shown in green. (a) uses a foreground model that fixes soil temperature in foreground model at 300\,K and reflection coefficient at 0.6, and (b) uses a model that fits the temperature and reflection coefficient as additional parameters with priors between 200-400\,K and 0.3-0.9 respectively.}
\label{fig:20percenthorizonheight}
\end{figure*}

We show that by fitting for \(T_\text{soil}\) and \(|\Gamma|\) with this unphysical prior range we are able to very consistently find the signal where the model that fixes values of \(T_\text{soil}\) and \(|\Gamma|\) is unable to.
This is exemplified in the case where we make \(S\) equal to 1.2, simulating an overestimation of the horizon in the foreground models of 20\%.
As detailed in Figure \ref{fig:20percenthorizonheight} we move from being entirely unable to recover the `True' signal with an RMSE of 0.0786 when we fix the parameters to those used in the mock data to a very accurate signal recovery.
Our fitted version has an RMSE approximately 9 times lower and a Log(\(\mathcal{Z}\)) 25 units higher.

By analysing the values of \(T_\text{soil}\) and \(|\Gamma|\) in Table \ref{tab:fittingtocorrect} we see how these parameters correct for horizon height error.
If the projected model of the horizon in our foreground correction is higher than the actual horizon the fitting process will compensate for this by dragging down the  soil temperature and increasing the magnitude of the reflection coefficient. 
This will by proxy increase the amount of sky that the telescope is `seeing' in comparison to the blackbody emission from the horizon.
This correction is useful, but not perfect, as a projected horizon height that is larger than the true height will mask out specific information on the sky power, obscuring parts of the spectral index map in our foreground maps, this is especially problematic when the galaxy is directly on the horizon.

If the model is lower than the actual height of the horizon the fitting method will do the opposite.
Here the model wants to maximise the amount of blackbody radiation coming from the horizon in order to compensate for the poor foreground modelling, and decrease the amount of sky reflection as much as possible to deal with the overestimation of the amount of sky we see.

While the Log(\(\mathcal{Z}\)) of our fits for an underestimation of the horizon are higher than when we systematically overestimate its height, we must be wary as the RMSE is also higher.
This would indicate that an underestimation of the horizon, even with fitting, will yield a worse recovery of the true signal.

\section{Conclusions and Future Work}
\label{sec:con}
This work aims to demonstrate the fact that a physically motivated foreground model demands an accurate description of the horizon for the recovery of the global 21\,cm signal.
We show that failing to account for the emissive and reflective properties of the soil on the horizon will lead to a non-detection of this signal.
This analysis was focused on REACH but is applicable to all global 21\,cm experiments that wish to use physically motivated foreground models in signal recovery.

This paper describes an easy-to-implement model that should greatly increase the ability of the REACH radiometer to recover the redshifted 21\,cm absorption trough in spite of the large mountainous horizon surrounding the antenna.

This model, while it is a great improvement in describing the horizon has a number of shortcomings that may be addressed in future work, these are as follows:

\begin{enumerate}
    
\item{The treatment of the horizon as composing entirely of one material that is entirely opaque to radio waves of all frequencies coming from behind is a bold one that must be addressed.
In a general context this is difficult, but a specific investigation of the REACH horizon may allow for discussion and mapping of vegetation, rocks and different kinds of soil, which will all have different dielectric permittivities, attenuating and reflecting radio waves with different strength.}

\item{This model also only discusses light rays, assuming that diffraction is negligible. 
Further studies may need to analyse this issue in more depth.
A proposed workaround to deal with diffraction would involve treating each spectral region as having a separate reflection coefficient, |\(\Gamma_i\)|, when reflected by the horizon which may be fit as an additional parameter.
This would allow for any region of the sky obscured by the model to be artificially scaled up again to compensate for the lack of explicit diffraction in the model.}

\item{This model treats the soil as having a constant temperature around the horizon, an improvement to this model would involve dividing the horizon into a number of regions based on cardinal direction.
Splitting these into a number of regions, each with the own temperature allows one to account for impact of the movement of the sun allowing for the eastern side of the valley to remain hotter than the western after the sun sets.}

\item{This model assumes an infinite, flat, ground describing no reflection or emission from the soil that falls below an altitude angle of 0\(^\circ\).
These near-field effects may have a very strong impact on signal recovery and will be explored in a later work}

\end{enumerate}

This work represents a large step forward in horizon modelling, demonstrating twofold decrease in RMSE from previous approaches to horizon modelling, with an increase in Log(\(\mathcal{Z}\)) \(\sim\) 1600.
We show that including a `hot' horizon is a necessity when trying to recover the 21\,cm signal using physically motivated foreground models.

We show that there is a dependency of signal recovery on the accurate estimation of soil parameters.
While there is some tolerance in the estimation of T\(_\text{soil}\), in which T\(_\text{soil}\) requires a precision of \(\approx\) 10\,K, \(|\Gamma|\) must be accurate to within 0.1 for accurate signal recovery.

To mitigate for error in soil parameter estimation we show that these parameters may be fitted for without compromising the integrity of signal recovery.
This fitting process will consistently perform as well as a horizon model in which the soil is described using free parameters that perfectly match those used in data generation. 

We also successfully demonstrate that allowing for these parameters to have priors that reach values beyond what is strictly expected physically will allow for a tolerance in horizon height measurement of up to \(\sim\) 20\%.

\section*{Acknowledgements}

We would like to thank Quentin Gueuning for providing the electromagnetic simulations of the log spiral antenna.
We would also like to thank Will Handley for his integral contributions to the REACH pipeline.

Joe Pattison, Dominic Anstey, and Eloy de Lera Acedo were supported by the Science and Technology Facilities Council.
We would also like to thank the Kavli Foundation for their support of REACH.

\section*{Data Availability}

The data that support the findings of this study are available from the first author upon reasonable request.

%%%%%%%%%%%%%%%%%%%%%%%%%%%%%%%%%%%%%%%%%%%%%%%%%% 



\bibliographystyle{mnras}
\bibliography{references} % if your bibtex file is called example.bib

\appendix

\section{Additional Tables}

\begin{table*}
\caption{Signal recovery by the REACH pipeline using a log spiral antenna of an injected Gaussian signal when \(T_\text{soil}\) and \(|\Gamma|\) in our foreground model are incorrectly chosen with respect to the mock data. \(T_\text{soil}\) in our data is set to be 300\,K and have \(|\Gamma|\) of 0.6. \(\mathcal{Z}_\text{Gaussian}\) is the Bayesian evidence of trying to fit for foregrounds with an additional Gaussian signal, and \(Z_\text{No 21}\) is the Bayesian evidence when we just fit for foregrounds.
\(\delta_{\text{Log}(\mathcal{Z}})\) is the difference in evidence between the models.
RMSE is the root mean squared error when comparing the `true' signal to one generated using the posterior averages generated by the Gaussian model.} 
\label{tab:changingtempandrefco}
\centering
\begin{tabular}{lllllllll}
\hline
&& F\(_0\) (MHz) & Width (MHz) & Depth (K) & Log(\(\mathcal{Z}_{\text{Gauss}}\)) & Log(\(\mathcal{Z}_{\text{No 21}}\)) & \(\delta_{\text{Log}(\mathcal{Z})}\) & RMSE\\
\hline
\hline
\hline
True Signal & & 85.0 & 15.0 & 0.155\\
\hline
\hline
T (K) & \(|\Gamma|\) \\
\hline
280 & 0.4 & 81.3\(\pm2.2\) & 17.6\(\pm1.5\) & 0.241\(\pm0.008\) & 197.4\(\pm0.4\) & 162.3\(\pm0.4\) & 35.1\(\pm0.6\) & 0.0471\\
 & 0.5 & 82.7\(\pm2.0\) & 17.5\(\pm1.3\) & 0.239\(\pm0.009\) & 234.0\(\pm0.4\) & 187.6\(\pm0.4\) & 46.4\(\pm0.6\) & 0.0449\\
 & 0.6 & 84.9\(\pm1.3\) & 16.8\(\pm1.2\) & 0.237\(\pm0.011\) & 270.1\(\pm0.4\) & 208.6\(\pm0.4\) & 61.5\(\pm0.6\) & 0.0408\\
 & 0.7 & 90.4\(\pm1.0\) & 12.6\(\pm1.1\) & 0.201\(\pm0.015\) & 285.7\(\pm0.4\) & 213.0\(\pm0.4\) & 72.7\(\pm0.6\) & 0.0245\\
 & 0.8 & 93.3\(\pm0.7\) & 10.6\(\pm0.6\) & 0.198\(\pm0.013\) & 261.4\(\pm0.4\) & 194.9\(\pm0.4\) & 66.5\(\pm0.6\) & 0.0339\\
 %\hline
 290 & 0.4 & 80.5\(\pm2.2\) & 17.8\(\pm1.4\) & 0.237\(\pm0.011\) & 222.9\(\pm0.4\) & 185.7\(\pm0.4\) & 37.2\(\pm0.6\) & 0.0468\\
 & 0.5 & 81.1\(\pm1.6\) & 18.2\(\pm1.2\) & 0.237\(\pm0.012\) & 261.8\(\pm0.4\) & 210.0\(\pm0.4\) & 51.8\(\pm0.6\) & 0.0475\\
 & 0.6 & 86.4\(\pm1.4\) & 14.7\(\pm1.4\) & 0.203\(\pm0.019\) & 287.1\(\pm0.4\) & 230.1\(\pm0.4\) & 57.0\(\pm0.6\) & 0.0200\\ 
 & 0.7 & 91.8\(\pm0.8\) & 10.9\(\pm0.7\) & 0.167\(\pm0.013\) & 281.0\(\pm0.4\) & 224.3\(\pm0.4\) & 56.7\(\pm0.6\) & 0.0268\\
 & 0.8 & 94.3\(\pm0.8\) & 10.5\(\pm0.5\) & 0.181\(\pm0.015\) & 235.4\(\pm0.4\) & 187.5\(\pm0.4\) & 47.9\(\pm0.6\) & 0.0351\\
 %\hline
300 & 0.4 & 78.3\(\pm2.2\) & 18.2\(\pm1.2\) & 0.232\(\pm0.015\) & 246.1\(\pm0.4\) & 208.8\(\pm0.4\) & 37.3\(\pm0.6\) & 0.0485\\
 & 0.5 & 81.3\(\pm1.9\) & 17.1\(\pm1.7\) & 0.220\(\pm0.021\) & 280.7\(\pm0.4\) & 234.8\(\pm0.4\) & 45.9\(\pm0.6\) & 0.0365\\
 & 0.6 & 88.9\(\pm1.3\) & 12.1\(\pm1.1\) & 0.154\(\pm0.016\) & 288.3\(\pm0.4\) & 244.4\(\pm0.4\) & 43.9\(\pm0.6\) & 0.0169\\ 
 & 0.7 & 92.8\(\pm1.0\) & 10.6\(\pm0.6\) & 0.146\(\pm0.014\) & 257.4\(\pm0.4\) & 221.0\(\pm0.4\) & 36.4\(\pm0.6\) & 0.0307\\
 & 0.8 & 95.2\(\pm1.0\) & 10.7\(\pm0.6\) & 0.168\(\pm0.017\) & 207.9\(\pm0.4\) & 175.2\(\pm0.4\) & 32.7\(\pm0.6\) & 0.0363\\
 %\hline
  310 & 0.4 & 77.5\(\pm2.5\) & 17.5\(\pm1.7\) & 0.232\(\pm0.015\) & 269.7\(\pm0.4\) & 234.6\(\pm0.4\) & 35.1\(\pm0.6\) & 0.0480\\
 & 0.5 & 84.2\(\pm1.6\) & 13.7\(\pm1.5\) & 0.167\(\pm0.021\) & 285.1\(\pm0.4\) & 255.6\(\pm0.4\) & 29.5\(\pm0.6\) & 0.0053\\
 & 0.6 & 90.6\(\pm1.3\) & 10.9\(\pm0.8\) & 0.121\(\pm0.015\) & 272.0\(\pm0.4\) & 246.0\(\pm0.4\) & 26.0\(\pm0.6\) & 0.0294\\ 
 & 0.7 & 94.0\(\pm1.1\) & 10.6\(\pm0.6\) & 0.128\(\pm0.017\) & 227.8\(\pm0.4\) & 207.7\(\pm0.4\) & 20.1\(\pm0.6\) & 0.0349\\
 & 0.8 & 96.3\(\pm1.5\) & 10.9\(\pm0.9\) & 0.160\(\pm0.020\) & 179.0\(\pm0.4\) & 157.9\(\pm0.4\) & 21.1\(\pm0.6\) & 0.0385\\
 %\hline
  320 & 0.4 & 77.5\(\pm2.5\) & 16.6\(\pm1.6\) & 0.208\(\pm0.022\) & 275.2\(\pm0.4\) & 249.3\(\pm0.4\) & 25.9\(\pm0.6\) & 0.0368\\
 & 0.5 & 86.0\(\pm1.8\) & 12.4\(\pm1.4\) & 0.118\(\pm0.018\) & 276.8\(\pm0.4\) & 259.1\(\pm0.4\) & 17.7\(\pm0.6\) & 0.0215\\
 & 0.6 & 141.0\(\pm48.9\) & 12.6\(\pm3.0\) & 0.071\(\pm0.030\) & 245.0\(\pm0.4\) & 232.8\(\pm0.4\) & 12.2\(\pm0.6\) & 0.0703\\ 
 & 0.7 & 181.8\(\pm6.5\) & 18.3\(\pm1.4\) & 0.063\(\pm0.009\) & 206.1\(\pm0.4\) & 186.4\(\pm0.4\) & 19.7\(\pm0.6\) & 0.0710\\
 & 0.8 & 176.0\(\pm5.9\) & 18.7\(\pm1.2\) & 0.088\(\pm0.012\) & 161.1\(\pm0.4\) & 138.3\(\pm0.4\) & 22.8\(\pm0.6\) & 0.0768\\
\hline
\end{tabular}
\end{table*}

\begin{table*}
\caption{An exploration into how fitting for temperature and reflection coefficient of soil surrounding the REACH antenna allows for the incorrect measurement of height of the horizon in the foreground models. We inject a signal with an 85M\,Hz Central Frequency, 15\,MHz Width, 0.155\,K Depth and our data model surrounds the telescope with soil at 300\,K with a \(|\Gamma|\) of 0.6. \(\mathcal{Z}_\text{Gaussian}\) is the Bayesian evidence of trying to fit the injected signal to a Gaussian, and \(\mathcal{Z}_\text{No 21}\) is the Bayesian evidence when we try to model for our data having no signal. \(\delta_{\text{Log}(\mathcal{Z})}\) is the difference in evidence between these models. RMSE refers to the root mean squared error when comparing the injected mock signal to one that we generate using the posterior averages that our Gaussian model suggests.}
\label{tab:fittingtocorrect}
\centering
\begin{tabular}{llllllllll}
\hline
& F\(_0\) (MHz) & Width (MHz) & Depth (K) & T(K) & \(|\Gamma|\) & Log(\(\mathcal{Z}_{\text{Gaussian}}\)) & Log(
\(\mathcal{Z}_{\text{No 21}}\)) & \(\delta_{\text{Log(Z)}}\) & RMSE\\
\hline

True & 85.0 & 15.0 & 0.155 & 300 & 0.60 \\

\hline
20 \% Over\\
No Fitting & 181.8\(\pm2.7\) & 19.5\(\pm0.5\) & 0.097\(\pm0.006\) &  &  & 249.9\(\pm0.4\) & 184.4\(\pm0.4\) & 65.5\(\pm0.6\) & 0.0786\\
Fitted & 86.5\(\pm2.0\) & 13.0\(\pm1.5\) & 0.174\(\pm0.023\) & 237.2\(\pm4.3\) & 0.80\(\pm0.03\) & 275.3\(\pm0.4\) & 249.2\(\pm0.4\) & 26.1\(\pm0.6\) & 0.0088\\
\hline
10 \% Over\\
No Fitting & 188.6\(\pm4.8\) & 18.1\(\pm1.6\) & 0.050\(\pm0.007\) &  &  & 264.2\(\pm0.4\) & 241.6\(\pm0.4\) & 22.7\(\pm0.6\) & 0.0685\\
Fitted & 87.3\(\pm1.9\) & 12.9\(\pm1.4\) & 0.174\(\pm0.024\) & 264.3\(\pm4.9\) & 0.71\(\pm0.03\) & 281.3\(\pm0.4\) & 252.4\(\pm0.4\) & 28.9\(\pm0.6\) & 0.0107\\
\hline
Correct \\ 
No Fitting & 88.9\(\pm1.3\) & 12.1\(\pm1.1\) & 0.154\(\pm0.016\) &  &  & 288.3\(\pm0.4\) & 243.5\(\pm0.4\) & 44.8\(\pm0.6\) & 0.0169\\
Fitted & 87.9\(\pm2.0\) & 13.1\(\pm1.5\) & 0.174\(\pm0.021\) & 294.9\(\pm5.3\) & 0.61\(\pm0.04\) & 287.0\(\pm0.4\) & 253.7\(\pm0.4\) &33.3\(\pm0.6\) & 0.0120\\
\hline
10 \% Under\\
No Fitting & 89.4\(\pm0.9\) & 14.3\(\pm1.1\) & 0.240\(\pm0.008\) &  &  & 267.3\(\pm0.4\) & 188.4\(\pm0.4\) & 78.9\(\pm0.6\) & 0.0375\\
Fitted & 89.2\(\pm1.9\) & 12.9\(\pm1.6\) & 0.170\(\pm0.020\) & 333.3\(\pm5.6\) & 0.48\(\pm0.04\) & 282.5\(\pm0.4\) & 253.3\(\pm0.4\) & 29.2\(\pm0.6\) & 0.0155\\
\hline
20 \% Under\\
No Fitting & 91.9\(\pm1.0\) & 13.1\(\pm1.2\) & 0.245\(\pm0.004\) &  &  & 205.0\(\pm0.4\) & 137.3\(\pm0.4\) & 67.7\(\pm0.6\) & 0.0419 \\
Fitted & 91.1\(\pm1.3\) & 11.9\(\pm1.3\) & 0.173\(\pm0.019\) & 377.6\(\pm5.3\) & 0.34\(\pm0.03\) & 279.5\(\pm0.4\) & 231.8\(\pm0.4\) & 47.7\(\pm0.6\) & 0.0229\\

\hline
\end{tabular}
\end{table*}

%\begin{thebibliography}{99}
%\end{thebibliography}

%%%%%%%%%%%%%%%%%%%%%%%%%%%%%%%%%%%%%%%%%%%%%%%%%%

%%%%%%%%%%%%%%%%% APPENDICES %%%%%%%%%%%%%%%%%%%%%

%%%%%%%%%%%%%%%%%%%%%%%%%%%%%%%%%%%%%%%%%%%%%%%%%%

% Don't change these lines
\bsp	% typesetting comment
\label{lastpage}
\end{document}